\documentclass[twocolumn,showpacs,preprintnumbers,amsmath,amssymb]{revtex4}

\usepackage{graphicx}
\usepackage{dcolumn}
\usepackage{bm}

\def\mr#1{\mathrm{#1}}

\DeclareMathOperator{\ineqr}{ }
\DeclareMathOperator{\ineql}{ }
\def\arangle{\stackrel{>}{\ineqr\limits_{\sim}}}
\def\alangle{\stackrel{<}{\ineql\limits_{\sim}}}
\makeatletter
\@addtoreset{footnote}{page}
\makeatother
 
\begin{document}

\preprint{}

\title{High Energy Neutrino Emission and Neutrino Background \\from Gamma-Ray Bursts in the Internal Shock Model}

\author{Kohta Murase}
 \email{kmurase@yukawa.kyoto-u.ac.jp}%
\author{Shigehiro Nagataki}%
 \email{nagataki@yukawa.kyoto-u.ac.jp}
\affiliation{%
Yukawa Institute for Theoretical Physics, Kyoto University,\\
Oiwake-cho, Kitashirakawa, Sakyo-ku, Kyoto, 606-8502, Japan
}%

\date{\today}
                        
\begin{abstract}
High energy neutrino emission from GRBs is discussed. In this paper, by using the simulation kit GEANT4, we calculate proton cooling efficiency including pion-multiplicity and proton-inelasticity in photomeson
production. First, we estimate the maximum energy of accelerated protons in GRBs. Using the obtained results, neutrino flux from one burst and a diffuse neutrino background  are evaluated quantitatively. We also take 
account of cooling processes of pion and muon, which are crucial for resulting neutrino spectra. We confirm the validity of analytic approximate treatments on GRB fiducial parameter sets, but also find that the effects of 
multiplicity and high-inelasticity can be important on both proton cooling and resulting spectra in some cases. Finally, assuming that the GRB rate traces the star formation rate, we obtain a diffuse neutrino background 
spectrum from GRBs for specific parameter sets. We introduce the nonthermal baryon-loading factor, rather than assume that GRBs are main sources of UHECRs. We find that the obtained neutrino background can be 
comparable with the prediction of Waxman \& Bahcall, although our ground in estimation is different from theirs. In this paper, we study on various parameters since there are many parameters in the model. The detection
 of high energy neutrinos from GRBs will be one of the strong evidences that protons are accelerated to very high energy in GRBs. Furthermore, the observations of a neutrino background has a possibility not only to test 
the internal shock model of GRBs but also to give us information about parameters in the model and whether GRBs are sources of UHECRs or not.
\end{abstract}

\maketitle

\section{\label{sec:level1}Introduction}
Gamma-ray bursts (GRBs) are the most powerful explosions in the universe. The observed isotropic energy can be estimated to be larger than 
${10}^{52}  \, \mr{ergs}$ \cite{Blo1,Fra1}. The high luminosity and the rapid time variability lead to the compactness problem. This problem and the 
hardness of observed photon spectra imply that $\gamma$-ray emission should be results of dissipation of kinetic energy of relativistic expanding shells. 
In the standard model of GRBs, such a dissipation is caused by internal shocks - internal collisions among the shells (see reviews e.g., 
\cite{Pir1,Zha1,Pir2}). Internal shock scenario requires a strong magnetic field, typically $10^{4} - 10^{7} \, \mr{G}$, which can accelerate electrons to 
high energy enough to explain observed $\gamma$-ray spectra by synchrotron radiation. Usually, Fermi acceleration mechanism is assumed to be working not 
only for electrons but also for protons, which can be accelerated up to a high energy within the fiducial GRB parameters. Physical conditions allow protons 
accelerated to greater than ${10}^{20} \, \mr{eV}$ and energetics can explain the observed flux of ultra-high-energy cosmic rays (UHECRs), assuming that 
similar energy goes into acceleration of electrons and protons in the shell \cite{Wax1,Vie1,Wax2,Bah1}. Such protons accelerated to the ultra-high energy 
cannot avoid interacting with GRB photons. This photomeson production process can generate very high energy neutrinos and gamma rays \cite{Wax3}. Whether 
GRBs are sources of UHECRs or not, internal shock models predict the flux of very high energy neutrinos at the Earth \cite{Der1,Gue1,Asa1}. Many authors 
have investigated neutrino emission from GRBs. Ice \v{C}herenkov detectors such as AMANDA at the South Pole have already been constructed and are taking 
data \cite{And2,Ahr1,Har1}. Now, the future $1\, {\mr{km}}^3$ detector, IceCube is being constructed \cite{Alv1,Ahr2,Ahr3}. In the Mediterranean Sea, 
ANTARES and NESTOR are under construction \cite{Kat1}. If the prediction is correct, these detectors may detect these neutrinos correlated with GRBs in 
the near future.\\
In this paper, we execute the Monte Carlo simulation kit GEANT4 \cite{Ago1} and simulate the photomeson production that causes proton energy loss. As a 
result, we can get meson spectrum and resulting neutrino spectrum from GRBs quantitatively. In our calculation, we take into account pion-multiplicity and 
proton-inelasticity, which are often neglected or approximated analytically in previous works, although they may be important in some cases 
\cite{Muc1,Muc2}. We also take into account the synchrotron loss of mesons and protons. These cooling processes play a crucial role for the resulting 
neutrino spectrum. Our models and method of calculation are explained in Sec. \ref{sec:level2}. One of our purposes is to seek physical conditions allowing 
proton to be accelerated up to ultra-high energy. Similar calculations are carried by Asano \cite{Asa1}, in which the possibility that a nucleon creates 
pions multiple times in the dynamical time scale is included but multiplicity is neglected. Using obtained results, we also investigate efficiency of 
neutrino emission for various parameter sets. Since there are many model uncertainties in GRBs, such a parameter survey is meaningful. Our final goal is 
to calculate a diffuse neutrino background from GRBs for specific parameter sets. A unified model for UHECRs from GRBs is very attractive 
\cite{Wax1,Vie1,Wax3}, although it requires GRBs being strongly baryon-loaded \cite{Wic1}. Even if observed UHECRs are not produced mainly by GRBs, high 
energy neutrino emission from GRBs can be expected. Hence, we leave the nonthermal baryon-loading factor as a parameter. In this paper, assuming GRBs trace 
star formation rate, we calculate a neutrino background from GRBs and compare our results with the flux obtained by Waxman \& Bahcall \cite{Wax2}. The design 
characters of neutrino detectors are being determined in part by the best available theoretical models, so numerical investigation of high energy neutrino 
fluxes should be important. Finally, we will consider the implications of neutrino observations and discuss a possibility that neutrino observation gives 
some information on the nonthermal baryon-loading factor and the inner engine, if the internal shock model is true. Our numerical results are shown in Sec. 
\ref{sec:level3}. Our summary and discussions are described in Sec. \ref{sec:level4}.

\section{\label{sec:level2}Method of Calculation}
\subsection{\label{subsec:levela}Physical Conditions}
Throughout this paper, we consider the epoch that internal shocks occurs (see reviews e.g., \cite{Pir1,Zha1,Pir2}). We focus on long GRBs, whose duration 
is typically $\sim 30 \, \mr{s}$ and they are likely to be related with the deaths of short-lived massive stars (see Sec. \ref{subsec:levele}). The internal
 shock model can produce the observed highly variable temporal profiles \cite{Kob1,Kob2}. In this model, the radial time scale is comparable with the 
angular time scale (see e.g., \cite{Pir1}), and each time scale is related to the pulse width. Widths of individual pulses vary in a wide range. GRB pulses 
with  $\sim 0.1 - 10 \, \mr{s}$ durations and separations for bright long bursts are typical ones \cite{Nak1,Nor1}. The shortest spikes have millisecond or 
even sub-millisecond widths in some bursts. It will reflect the intermittent nature of the fireball central engine. The internal collision radii are 
determined by the separation of the subshells, which is written by $d$ in the comoving frame. Since the observed pulse width is given by $\delta t \approx 
d/2\Gamma c$, the internal collision radius is expressed by the commonly used relation, $r \approx 2\Gamma ^2 c \delta t \approx 6 \times 10^{14} 
{(\Gamma/300)}^{2}(\delta t/0.1 \, \mr{s}) \, \mr{cm}$, where $\Gamma$ is a bulk Lorentz factor of the shell, which is larger than 100. We consider 
collision radii in the range of $(10^{13} - 10^{16}) \, \mr{cm}$ in our calculation. The width of the subshells is constrained by the observed variability 
time and the duration, and it is typically written by $l \approx r/\Gamma$, where $r$ is the radius at which internal shocks begin to occur. In this paper, we leave $l$ as a 
parameter because the precise width of each subshell is unknown. The dynamical time of each collision is $t_{dyn} \approx l/c$. For simplicity, we assume
each collision radiates similar energy, $E_{\gamma}^{iso} = E_{\gamma,tot}^{iso}/N$, where $N$ is the number of collisions and it is almost the number of 
subshells \cite{Kob1}. Although we do not know about $N$ precisely, we set $N \sim (10-100)$ in our calculation since the number of pulses per bright long 
burst is typically the order of dozens \cite{Mit1}.\\
The geometrically corrected GRB radiation energy is typically $E_{\gamma,tot}=1.24 \times 10^{51} h_{70}^{-1} \, \mr{ergs}$ \cite{Blo1,Fra1}, and we fix 
$E_{\gamma,tot}$ throughout this paper. The isotropic energy is a few orders of magnitude larger than this energy, and we take 
$E_{\gamma,tot}^{iso}=f_b^{-1} E_{\gamma,tot} \sim (10^{52}-10^{54}) \, \mr{ergs}$. Here, $f_{b} \equiv \theta _{j}^{2}/2$ is a beaming factor of GRB jet. 
In the prompt phase $1/\Gamma \alangle \theta _{j}$ has to be satisfied. Once we determine $E_{\gamma}$ and $N$, $f_b$ is also determined.\\
The photon energy density in the comoving frame is given by,
\begin{equation}
U_{\gamma}=\frac{E_{\gamma}^{iso}}{4\pi \Gamma r^2 l} \label{phden}
\end{equation}
In the standard model, the prompt spectrum is explained by the synchrotron radiation from electrons. For simplicity, we assume that the GRB photon spectrum 
obeys a power-law spectrum. That is,
\begin{equation}
\frac{dn}{d\varepsilon} \propto \left\{ \begin{array}{rl} {\varepsilon}^{-\alpha} & \mbox{(for $\varepsilon ^{\mr{min}} < \varepsilon 
< \varepsilon ^b$)}\\
                                                  {\varepsilon}^{-\beta} & \mbox{(for $\varepsilon ^b < \varepsilon < \varepsilon ^{\mr{max}}$)} 
\end{array} \right.
\end{equation}
The observed break energy is $\sim 250 \, \mr{keV}$, which corresponds to the break energy in the comoving frame, $\varepsilon ^{b} \sim$ a few $\mr{keV}$. 
So we set $\varepsilon ^{b}$ to 1 keV for $E_{\gamma}^{iso}=2 \times 10^{51} \, \mr{ergs}$, 3 keV for $E_{\gamma}^{iso}=2 \times 10^{52} \, \mr{ergs}$, 6 
keV for $E_{\gamma}^{iso}=2 \times 10^{53} \, \mr{ergs}$. In addition, we set spectral indices to $\alpha = 1$, $\beta = 2.2$ as fiducial values and 
$\alpha = 0.5$, $\beta = 1.5$ as a flatter case. We take the minimum energy as $1 \, \mr{eV}$ and the maximum energy as $10 \, \mr{MeV}$. Below 
$1 \, \mr{eV}$, the synchrotron self-absorption will be crucial, while above $10 \, \mr{MeV}$, the pair creation will be important \cite{Asa2}.\\ 
To explain the observed emission, the internal shock model requires the strong magnetic energy density, which is expressed by 
$U_{B}={\epsilon}_{B}U_{\gamma}$. We take ${\epsilon}_{B}=0.1, 1, 10$ as a parameter. 

\subsection{\label{subsec:levelb}Acceleration and Cooling of Proton}
The radiating particles are considered to gain their energies by stochastic acceleration. The most promising acceleration process is Fermi 
acceleration mechanism. We assume that electrons and protons will be accelerated in GRBs by first order Fermi-acceleration by diffusive scattering of 
particles across storng shock waves. The acceleration time scale is given by $t_{acc} \approx (3cr_{L}/{{\beta}_{A}}^{2}){(B/\delta {B})}^{2}$, where 
$r_L$ is Larmor radius of the particle, ${\beta}_{A}=v_{A}/c$, $v_{A}$ is Alfv\'en velocity, and $\delta B$ is the strength of the turbulence in the 
magnetic field \cite{Kul1,Rac1}. Assuming that the diffusion coefficient is proportional to the Bohm diffusion coefficient, we can write the acceleration 
time of proton as follows,
\begin{equation}
t_{acc}\equiv \eta \frac{r_{L}}{c}=\eta\frac{\varepsilon _{p}}{eBc} \label{acc}
\end{equation}
In usual cases $\eta \arangle 10$ would be more realistic, and $\eta \sim 1$ will give a reasonable lower limit for any kind of Fermi acceleration time 
scale. As for the ultrarelativistic shock acceleration, the acceleration time scale can be written $t_{acc} \sim \varepsilon_{p}/\sqrt{2} \Gamma _{rel} 
eBc$, where $ \Gamma _{rel}$ is the relative Lorentz factor of the upstream relative to the downstream \cite{Gal1}. For the case of mildly relativistic 
shocks, $\eta \sim 1$ is probably possible \cite{Wax4}. Hence, we set $\eta =1$ optimistically in our calculation. \\
Corresponding to $t_{acc}<t_{dyn}$, proton energy is restricted by $\varepsilon _{p} \alangle eBl/\eta$, which satisfies the requirement that proton's 
Larmor radius has to be smaller than the size of the acceleration sites. Even when GRBs are jet-like, the transverse size will be larger than the radial 
size $r/\Gamma$ because the jet opening angle $\theta_{j}$ is larger than $1/\Gamma$, so $t_{acc}<t_{dyn}$ holds. Proton's maximal energy is also 
constrained by various cooling processes, for example, proton synchrotron cooling, inverse Compton (IC) cooling, and photohadronic cooling and so on. 
First, the synchrotron loss time scale for relativistic protons is,
\begin{equation}
t_{syn}=\frac{3m_{p}^4 c^3}{4 \sigma _{T} m_{e}^2}\frac{1}{\varepsilon_{p}}\frac{1}{U_{B}} \label{sync}
\end{equation} 
Second, we have to consider IC cooling, but we ignore IC process in this paper for simplicity. This contribution is considered to be small in the region 
where proton energy is above around $10^{17} \mr{eV}$ \cite{Asa1}. Although IC will be more important than synchrotron loss when proton energy is below 
that energy, we can ignore this fact because photohadronic loss will dominate IC for our parameters. Furthermore, in our cases IC does not matter for 
the purpose to determine the proton maximum energy. Third, the photohadronic cooling process is important, and pions and resulting neutrinos appear through 
pion decays. When the internal shock occurs at small radii, photohadronic cooling may dominate synchrotron cooling. We  will discuss this process 
in the next subsection. Finally, we take into account the adiabatic cooling process, which has a time scale $t_{ad}$ independent of the proton energy, and 
direct ejection of protons from the emission region. The latter may be dependent on the proton energy if diffusive losses are relevant. For simplicity, 
we neglect diffusive losses and assume that protons are confined over the time scale set by adiabatic expansion. From above, the total proton loss time 
scale is expressed as follows,
\begin{equation}
t_{p}^{-1} \equiv t_{p\gamma}^{-1} + t_{syn}^{-1} + t_{ad}^{-1} \le  t_{acc}^{-1} \label{mac}
\end{equation}

The maximum energy of proton is determined by this inequality, but we do not know the minimum energy. We set the minimum energy of proton to $10 \, 
\mr{GeV}$. Proton energy density can be expressed by $U_{p}=\epsilon_{acc}U_{\gamma}$, where $\epsilon_{acc}$ is a nonthermal baryon-loading factor. 
However, we have few knowledge about how many protons can be accelerated for now. So, we adopt $\epsilon _{acc} = 10$ as a fiducial parameter, assuming 
that a significant fraction of relativistic protons can be accelerated. It is also assumed that proton distribution has $dn_{p}/d \varepsilon _{p} \propto 
\varepsilon _{p}^{-2}$ according to first order Fermi acceleration mechanism.

\subsection{\label{subsec:levelc}Photomeson Production}
GRB photons produced by synchrotron radiation and relativistic protons accelerated in the internal shocks can make mesons such as ${\pi}^{\pm }$ and 
${\mr{K}}^{\pm }$, and secondary electrons, positrons, gamma rays, and neutrinos through the photomeson production process. Some authors used 
$\Delta$-approximation for simplicity, but this approximation underestimates proton energy loss in the high energy region. This is because the theoretical 
model predicts the increase of cross section in the high energy regions. We use the Monte Carlo simulation toolkit GEANT4 \cite{Ago1}, which includes cross 
sections up to 40 TeV \cite{Kos1}. Above 40 TeV, we extrapolate cross sections up to the maximum energy, which is around $10 \, \mr{PeV}$. We checked the 
accuracy of GEANT4 photomeson cross section within a few percent, comparing with PDG data \cite{PDG1}. Moreover, we take into account inelasticity. Around 
the $\Delta $-resonance, inelasticity is about $0.2$. But above the resonance, inelasticity increases with energy, and the value is about $0.5-0.7$, that 
is consistent with M\"ucke et al. \cite{Muc1}. However, it seems that the current version of GEANT4 tends to overestimate the amount of produced neutral 
pions, leading to underestimation of the amount of charged pions by a factor of $\sim 3$. So we approximate ${\pi}^{+}:{\pi}^{0}=1:1$ for single-pion 
production, and ${\pi}^{+}{\pi}^{-}:{\pi}^{+}{\pi}^{0}=7:4$ for double-pion production \cite{Sch1,Rac1}.\\
Photomeson cooling time scale of relativistic protons is given by the following equation for isotropic photon distribution \cite{Wax3}.
\begin{equation}
t^{-1}_{p\gamma}(\varepsilon _{p}) = \frac{c}{2{\gamma}^{2}_{p}} \int_{\bar{\varepsilon}_{th}}^{\infty} \! \! \! d\bar{\varepsilon} \, 
{\sigma}_{p\gamma}(\bar{\varepsilon}) {\kappa}_{p}(\bar{\varepsilon})
\bar{\varepsilon} \int_{\bar{\varepsilon}/2{\gamma}_{p}}^{\infty} \! \! \! \! \! \! \! \! \! d \varepsilon \, {\varepsilon}^{-2} \label{pcool}
\frac{dn}{d\varepsilon}
\end{equation}
where $\bar{\varepsilon}$ is the photon energy in the rest frame of proton, $\gamma _{p}$ is the proton's Lorentz factor, $\kappa _{p}$ is the inelasticity
of proton, and $\bar{\varepsilon} _{th}$ is the threshold photon energy for photomeson production in the rest frame of the incident proton, which is 
$\bar{\varepsilon}_{th} \approx 145 \, \mr{MeV}$. Using this equation, we can obtain photomeson production efficiency. Protons with $1 \, \mr{PeV}$ energy 
will effectively interact with soft X-ray photons with energy $\arangle 0.16 \mr{keV}$.\\
Pion spectrum can be obtained by executing GEANT4 through the following equation, 
\begin{equation}
\frac{dn_{\pi}}{d\varepsilon _{\pi}}=\int _{\varepsilon _{p}^{\mr{min}}}^{\varepsilon _{p}^{\mr{max}}} \!\!\!\!\!  d \varepsilon _{p} \, \frac{d n_{p}}
{d \varepsilon _{p}} \int _{\varepsilon ^{\mr{min}}}^{\varepsilon ^{\mr{max}}} \!\!\!\!\! d \varepsilon \, \frac{dn}{d\varepsilon}
\int \frac{d \Omega}{4\pi} \, \frac{d \sigma _{p\gamma}(\varepsilon,\Omega,\varepsilon _{p}) \xi}{d\varepsilon _{\pi}} \, c \, t_{p} \label{getpion}
\end{equation}
where $d n_{p}/d \varepsilon _{p}$ and $d n/d \varepsilon$ is proton and photon distribution in the comoving frame, $\xi$ is the pion-multiplicity, and 
$t_{p}$ is the proton loss time scale, which is defined by the equation (\ref{mac}). 

\subsection{\label{subsec:leveld}Neutrinos from Pion and Muon Decay}
By using GEANT4, we can get pion spectra. Hence, neutrino spectra follow from the spectra of pions and muons. Neutrinos are produced by the decay 
of ${\pi}^{\pm} \rightarrow {\mu}^{\pm}+{\nu}_{\mu}({\bar{\nu}}_{\mu}) \rightarrow e^{\pm}+{\nu}_{e}({\bar{\nu}}_{e})+{\nu}_{\mu}+{\bar{\nu}}_{\mu}$. 
The life times of pions and muons are $t_{\pi}={\gamma}_{\pi} {\tau}_{\pi}$ and   $t_{\mu}={\gamma}_{\mu} {\tau}_{\mu}$ respectively. Here, ${\tau}_{\pi}
=2.6033 \times 10^{-8} \mr{s}$ and ${\tau}_{\mu}=2.1970 \times 10^{-6} \mr{s}$ are the mean life times of each particle. 
When pions decay with the spectrum, $dn_{\pi}/d\varepsilon _{\pi}$ by ${\pi}^{\pm} \rightarrow {\mu}^{\pm}+{\nu}_{\mu}({\bar{\nu}}_{\mu})$, the spectrum 
of neutrinos are given by \cite{Der2},
\begin{equation}
\frac{dn_{\nu}}{d\varepsilon _{\nu}}=\frac{m_{\pi}c}{2\varepsilon _{\nu}^{*}} \int ^{\infty}_{\varepsilon _{\pi}^{\mr{min}}} \! \! \! \! \!
d\varepsilon _{\pi} \, \frac{1}{p_{\pi}} \frac{dn_{\pi}}{d\varepsilon _{\pi}}
\end{equation}
Here, $\varepsilon_{\nu}^{*}=(m_{\pi}^2-m_{\mu}^2)c^2/2m_{\pi}$ and $\varepsilon _{\pi}^{\mr{min}}=(\varepsilon_{\nu}^{*}/\varepsilon _{\nu}+\varepsilon
 _{\nu}/\varepsilon_{\nu}^{*})/2$.
Similarly, we can get muon spectrum from pion spectrum. Muon decay is the three-body-decay process, which is slightly more complicated than the case of
 two-body-decay. Given the spectrum of muon, it can be calculated by the following equation \cite{Sch2}, 
\begin{equation}
\frac{dn_{\nu}}{d\varepsilon _{\nu}}=\int _{m_{\mu}c^2}^{\infty} \! \! \! \! \! d\varepsilon _{\mu} \, \frac{1}{cp_{\mu}} \frac{d n_{\mu}}
{d\varepsilon _{\mu}} \int _{\varepsilon _{\mu 1}^{*}}^{\varepsilon _{\mu 2}^{*}} \! \! \! d \varepsilon _{\nu}^{*} \, \frac{1}{\varepsilon _{\nu}^{*}} 
(f_0(\varepsilon _{\nu}^{*}) \mp \mr{cos} \theta _{\nu}^{*} f_1(\varepsilon _{\nu}^{*}))
\end{equation}
where $\varepsilon _{\mu 1}^{*}=\gamma _{\mu}\varepsilon _{\nu} - {(\gamma _{\mu}^{2}-1)}^{1/2}\varepsilon _{\nu}$, $\varepsilon _{\mu 2}^{*}=\mr{min}
(\gamma _{\mu}\varepsilon _{\nu} + {(\gamma _{\mu}^{2}-1)}^{1/2}\varepsilon _{\nu}, \, (m_{\mu}^2-m_{e}^2)c^2/2m_{\mu})$, and $\varepsilon _{\nu}^{*}$ 
is the energy of muon-neutrino in the muon-rest frame. $f_0(x)$ and $f_1(x)$ can be calculated by the quantum field theory, which are, $f_0(x)=2x^2(3-2x)$, 
$f_1 (x)=2 x^2 (1-2x)$, and $x=2\varepsilon _{\nu}^{*} / m_{\mu} c^2$. However, because of cooling processes of ${\pi}^{\pm}$ and ${\mu}^{\pm}$, these 
equations have to be applied at each time step. The pion or muon decay probability is $(1-\mr{exp}(-\Delta t/t _{\pi,\mu}))$, and meanwhile cooled by 
synchrotron cooling and adiabatic cooling. The synchrotron radiating time scale is given by replacing proton mass with pion or muon mass in the equation 
(\ref{sync}). The adiabatic cooling time scale is still comparable to dynamical time scale. We neglect IC process of pions and muons in our calculation,
 because Klein-Nishina suppression will work in our cases \cite{Asa1}. We also neglect neutrinos due to neutron decay 
$n \rightarrow p+e^{-}+{\bar{\nu}}_{e}$, whose time scale is much larger than the dynamical time scale $t_{dyn}$.\\
Roughly speaking, in the $\Delta$-resonance picture, neutrino spectrum follows proton spectrum above the break energy which is given by, 
$\varepsilon _{p}^{b} \varepsilon ^{b} \sim 0.3 \, \mr{GeV}$, and becomes harder below the break by $\varepsilon _{p}^{\beta -1}$ due to $f_{p\gamma} 
\equiv t_{dyn}/t_{p\gamma} \propto \varepsilon _{p}^{\beta-1}$ \cite{Wax3}. In addition, if $t_{\pi,\mu} > t_{ad}$, neutrino spectrum will be suppressed 
by $t_{ad}/t_{\pi,\mu}$ due to adiabatic cooling. If $t_{\pi,\mu} > t_{syn}$, neutrino spectrum will be suppressed by $t_{syn}/t_{\pi,\mu}$ due to 
synchrotron cooling. These statements will be confirmed numerically by above methods.

\subsection{\label{subsec:levele}GRB Diffuse Neutrino Background}
UHECRs may come from GRBs \cite{Wax1,Vie1}. This hypothesis requires that the conversion of the initial energy of a fireball into UHECRs must be very 
efficient. In other words, this needs a very large value of $\epsilon _{acc}$. On the other hand, Waxman \& Bahcall \cite{Wax2} have derived a model 
independent upper bound for high energy neutrino background from sources optically thin to photomeson production. This bound is conservative and robust 
\cite{Bah1} and it is called as the Waxman and Bahcall limit (WB limit). They have also estimated the contribution of GRBs to the high energy neutrino 
background, assuming that UHECRs come from GRBs, and showed this contribution is consistently below the WB limit.\\
Whether UHECRs come from GRBs or not, numerical calculations on neutrino spectrum obtained by our method shown above can provide a diffuse neutrino 
background quantitatively, assuming the GRB rate follows some distribution. The demonstration that long-duration GRBs are associated with core-collapse 
supernovae \cite{Gal2} implies that GRB traces the deaths of short-lived massive stars. Furthermore, GRBs can be detected to very high redshifts, 
unhindered by intervening dust and the current record is recently observed GRB 050904 \cite{Pri1}. This holds the promise of being useful tracers of star 
formation rate (SFR) in the universe \cite{Tot1,Wij1,Pac1}. However, there remain many problems such as observational bias and the effect of 
dust-enshrouded infrared starbursts, although these are crucial for to which extend GRBs follow the SFR and to which extend they can be used to determine 
the SFR at high redshifts.\\
In this paper, we estimate neutrino background for several models, following Nagataki et al. \cite{Nag1,And1}. First, we adopt the hypothesis that GRB rate
 traces the SFR, $R_{\mr{GRB}} \propto R_{\mr{SF}}$. This ansatz that GRBs are likely to trace the observed SFR in a globally averaged sense, is not 
precluded for now although there are some uncertainties as described above. Second, we use parameterization of Porciani \& Madau \cite{Por1}, especially 
employing their models SF1, SF2 and SF3, which are expressed by following equation in a flat Friedmann-Robertson-Walker universe,
\begin{subequations}
\begin{eqnarray}
{\psi}_{*}^{\mr{SF1}}(z)&=&0.32f_{*}h_{70}\frac{\mr{exp}(3.4z)}{\mr{exp}(3.8z)+45} \nonumber \\
&& \times F(z,\Omega _{m},\Omega _{\Lambda})M_{\odot} \, \, \, {\mr{yr}}^{-1}{\mr{Mpc}}^{-3} \label{SF1}
\end{eqnarray}
\begin{eqnarray}
{\psi}_{*}^{\mr{SF2}}(z)&=&0.16f_{*}h_{70}\frac{\mr{exp}(3.4z)}{\mr{exp}(3.4z)+22} \nonumber \\
&& \times F(z,\Omega _{m},\Omega _{\Lambda})M_{\odot} \, \, \, {\mr{yr}}^{-1}{\mr{Mpc}}^{-3} \label{SF2}
\end{eqnarray}
\begin{eqnarray}
{\psi}_{*}^{\mr{SF3}}(z)&=&0.21f_{*}h_{70}\frac{\mr{exp}(3.05z-0.4)}{\mr{exp}(2.93z)+15}  \nonumber \\
&& \times F(z,\Omega _{m},\Omega _{\Lambda})M_{\odot} \, \, \, {\mr{yr}}^{-1}{\mr{Mpc}}^{-3} \label{SF3}
\end{eqnarray}
\end{subequations}
where $F(z,\Omega _{m},\Omega _{\Lambda})={(\Omega_{m}{(1+z)}^{3}+\Omega _{\Lambda})}^{1/2}/{(1+z)}^{3/2}$, $h_{70}=H_0/70 \, \mr{km} \, {\mr{s}}^{-1} \, 
{\mr{Mpc}}^{-1}$, and we adopt the standard $\Lambda$CDM cosmology $(\Omega _m=0.3, \Omega _{\Lambda}=0.7)$.
The correction factor $f_{*}$ is introduced for uncertainties of SFR. We set $f_{*}=1$, which is consistent with mildly dust-corrected UV data at low 
redshift; on the other hand, it may underestimate the results of other wave band observations and we only have to correct $f_{*}$ in such cases \cite{Bal1}.
 Many works have modeled the expected evolution of the cosmic SFR with redshift, but there are some uncertainties (in particular at high redshift z 
$\arangle 6$). In addition, even observational estimates at modest high redshift have been plagued by uncertainties arising from a result of correction for 
dust extinction. Due to a few constraints at high redshift, we cannot avoid extrapolating SFRs. For these reasons, we employ three models.\\
The inner engine of GRBs is still unknown, although there are several plausible candidates. One of the plausible models of GRB progenitors is a collapsar 
model \cite{Mac1}, because GRBs likely have a link with the explosion of a massive ($M \arangle (35-40) \, M_{\odot}$) rotating star whose core collapses to 
form a black hole \cite{Mac1,Fry1,Pet1}. Assuming that massive stars with masses larger than $\sim 35 \, M_{\odot}$ explode as GRBs, GRB rate can be 
estimated for a selected SFR by multiplying the coefficient,
\begin{equation}
f_{cl}\times\frac{\int_{35}^{125}dm \, \phi(m)}{\int_{0.4}^{125}dm \, m\phi(m)}=1.5 \times 10^{-3} f_{cl} M_{\odot}^{-1} \label{IMF}
\end{equation}
where $\phi (m)$ is the initial mass function (IMF) and mass is the stellar mass in solar units. Here we adopt the Salpeter's IMF ($\phi(m) \propto 
m^{-2.35}$), assuming that the IMF does not change with time, which may be a good approximation if there are no significant correlations between the 
IMF and the environment in which stars are born. For now, extant evidences seem to argue against such correlations at $z \alangle 2$, although this 
validity at high redshift is uncertain \cite{Sca1}. For comparison, we can obtain $R_{\mr{SN}}(z)=0.0122M_{\odot}^{-1}{\psi}_{*}^{\mr{SF}}(z)$ assuming 
that all stars with $M>8M_{\odot}$ explode as core-collapse supernovae. This result combined with $f_{*}=1$ agrees with the observed value of local 
supernova rate, $R_{\mr{SN}}(0)=(1.2\pm 0.4)\times 10^{-4} h_{70}^{3}\, {\mr{yr}}^{-1}{\mr{Mpc}}^{-3}$ \cite{Mad1}. In the equation (\ref{IMF}), we 
introduce an unknown parameter, $f_{cl}$, which expresses the fraction of the collapsars whose mass range is in $(35-125)M_{\odot}$ are accompanied with 
GRBs. We normalize $f_{cl}$ using the value of GRB rate, $R_{\mr{GRB}}(0)=17h_{70}^{3} \, {\mr{yr}}^{-1}{\mr{Gpc}}^{-3}$, obtained by recent analysis using 
BATSE peak flux distribution \cite{Gue2,Sch3}. From equations (\ref{SF1}), (\ref{SF2}), and (\ref{SF3}), combined with (\ref{IMF}), we obtain following 
expressions in units of ${\mr{yr}}^{-1}{\mr{Gpc}}^{-3}$,
\begin{subequations}
\begin{eqnarray}
R_{\mr{GRB1}}(z)&=&17 \left( \frac{f_{cl}}{1.6\times 10^{-3}}\right) \frac{46 \, \mr{exp}(3.4z)}{\mr{exp}(3.8z)+45} \, \, \, \, \, \, \, \, \, \, \, \, \, \,
 \,   \, \, \, \, \, 
\end{eqnarray}
\begin{eqnarray}
R_{\mr{GRB2}}(z)&=&17 \left( \frac{f_{cl}}{1.6\times 10^{-3}}\right) \frac{23 \, \mr{exp}(3.4z)}{\mr{exp}(3.4z)+22} \, \, \, \, \, \, \, \, \, \, \, \, \, \,
 \,   \, \, \, \, \, 
\end{eqnarray}
\begin{eqnarray}
R_{\mr{GRB3}}(z)&=&21 \left( \frac{f_{cl}}{1.6\times 10^{-3}} \right) \frac{24 \, \mr{exp}(3.05z-0.4)}{\mr{exp}(2.93z)+15}  \, \, \, \, \,  \, \, \, \, \, 
\end{eqnarray}
\end{subequations}
We also consider the Rowan-Robinson SFR \cite{Row1} that can be fitted with the expression in units of ${\mr{yr}}^{-1}{\mr{Gpc}}^{-3}$,
\begin{eqnarray}
R_{\mr{RR}}(z)=41 \left( \frac{f_{cl}}{1.6\times 10^{-3}} \right) \left\{\begin{array}{rl} &10^{0.75z} \, \mbox{(for z $<$ 1)} \\
                                                  &10^{0.75} \, \, \, \mbox{(for z $>$ 1)} 
\end{array} \right.
\end{eqnarray}
The parameter, $f_{cl}$ is unknown for now, but $\sim 2 \times 10^{-3}$ will give the reasonable upper limit for the probability for one collapsar to 
generate a GRB, which corresponds to the value if all the GRBs come from collapsars. On the other hand, direct estimates from the sample of GRBs with 
determined redshifts are contaminated by observational biases and are insufficient to determine the precise rate and luminosity function. In addition, the 
observed peak luminosity also depends on intrinsic spectrum. More precise data on GRBs will give us information on the accurate GRB rate. For these reasons,
 we leave the value of $f_{cl}$ as a parameter here.\\ 
In order to get the differential number flux of background neutrinos, first we compute the present number density of the background neutrinos per unit 
energy from GRBs. The contribution of neutrinos emitted in the interval of the redshift $z \sim z+dz$ is given as,
\begin{equation}
dn_{\nu}^{\mr{ob}}(E_{\nu})=R_{\mr{GRB}}(z){(1+z)}^{3}\frac{dt}{dz}dz\frac{dN_{\nu}(E_{\nu}^{'})}{dE_{\nu}^{'}}dE_{\nu}^{'}{(1+z)}^{-3}
\end{equation}
where $E_{\nu}^{'}=(1+z)E_{\nu}$ is the energy of neutrinos at redshift z, which is now observed as $E_{\nu}$ and $dN_{\nu}(E_{\nu}^{'})/dE_{\nu}^{'}$ is 
the number spectrum of neutrinos emitted by one GRB explosion. The differential number flux of GRB background neutrinos, $dF_{\nu}/dE_{\nu}$, using the 
relation $dF_{\nu}/dE_{\nu}=c \, dn_{\nu}^{\mr{ob}}/dE_{\nu}$,
\begin{eqnarray}
\frac{dF_{\nu}}{dE_{\nu}d\Omega}&=&\frac{c}{4\pi H_{0}} \int _{z_{\mr{min}}}^{z_{\mr{max}}} dz \, R_{\mr{GRB}}(z) \frac{dN_{\nu}((1+z)E_{\nu})}
{dE_{\nu}^{'}}  \nonumber \\ && \times \frac{1}{\sqrt{(1+\Omega_{m}z){(1+z)}^{2}-\Omega_{\Lambda}(2z+z^2)}}
\end{eqnarray}
where we assume $z_{\mr{min}}=0$, and $z_{\mr{max}}=7$ or $z_{\mr{max}}=20$. This is because the high redshift GRB event GRB 050904 is recently reported by 
Swift observation and the GRB distribution likely extends beyond $z=6$ \cite{Pri1}. In addition, some GRBs are expected to exist at much higher
redshifts than $z=7$. First, preliminary polarization data on the cosmic microwave background collected by WMAP indicate a high electron scattering 
optical depth, hinting that the first stellar objects in the universe should have formed as early as $z \sim 20$ \cite{Spe1}. Second, theoretical 
simulations of the formation of the first stars similarly conclude that these should have formed at redshifts $z \sim (15-40)$ \cite{Abe1}. Because there 
are convincing evidences that at least long GRBs are associated with the deaths of massive stars, it is conceivable that high-z GRBs ($z \arangle 15-20$ or 
even higher) exist.\\
\begin{figure*}[tb]
\begin{minipage}{.48\linewidth}
\includegraphics[width=\linewidth]{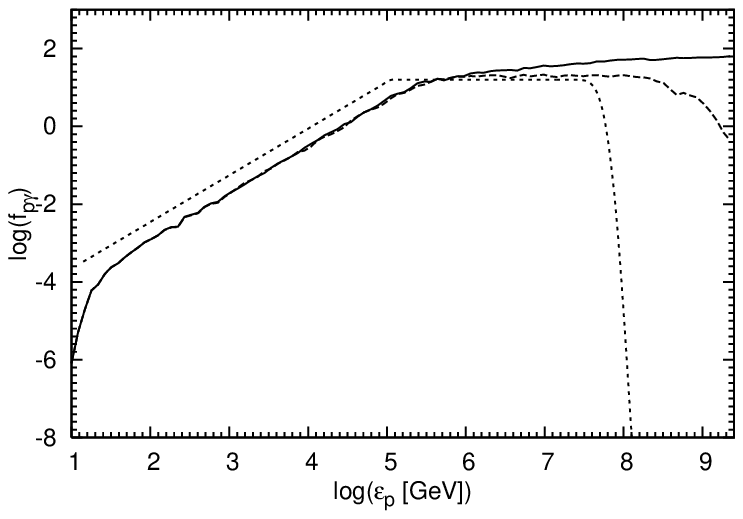}
\caption{\label{PC1} Proton cooling efficiencies, $f_{p\gamma}$, for $r=2 \times {10}^{13} \, \mr{cm}$ and $E_{\gamma}^{iso}=2 \times {10}^{51} \, 
\mr{ergs}$, by GEANT4 (solid line). For comparison, we also show the case with cross section having a cutoff at 2 GeV (dashed line) and rough analytic 
approximation (dotted line). }
\end{minipage}
\begin{minipage}{.02\linewidth}
\includegraphics[width=\linewidth]{white.eps}
\end{minipage}
\begin{minipage}{.48\linewidth}
\includegraphics[width=\linewidth]{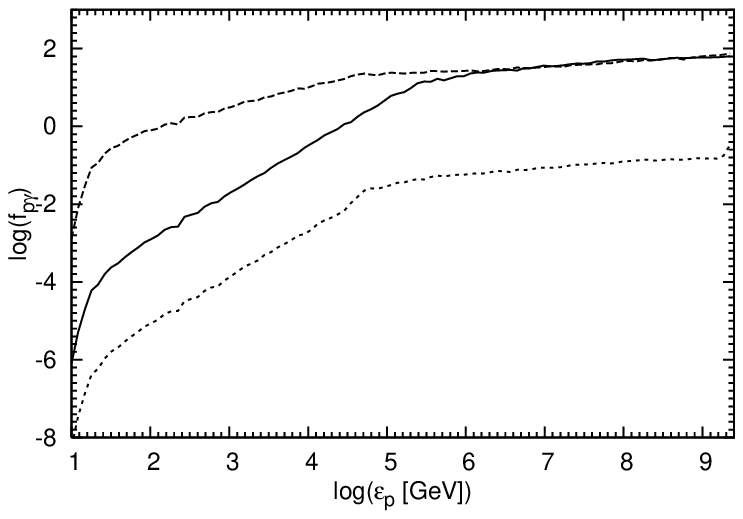}
\caption{\label{PC2} Proton cooling efficiencies, $f_{p\gamma}$; case A for $r=2 \times {10}^{13} \, \mr{cm}$ and $E_{\gamma}^{iso}=2 \times {10}^{51} \, 
\mr{ergs}$ (solid line), case B for $r=5.4 \times {10}^{14} \, \mr{cm}$ and $E_{\gamma}^{iso}=2 \times {10}^{52} \, \mr{ergs}$ (dotted line), case C for 
$r=2 \times {10}^{13} \, \mr{cm}$ and $E_{\gamma}^{iso}=2 \times {10}^{53} \, \mr{ergs}$ (dashed line). Each shell width is given by $l=r/\Gamma$.}
\end{minipage}
\end{figure*}
\begin{figure*}[bt]
\begin{minipage}{.48\linewidth}
\includegraphics[width=\linewidth]{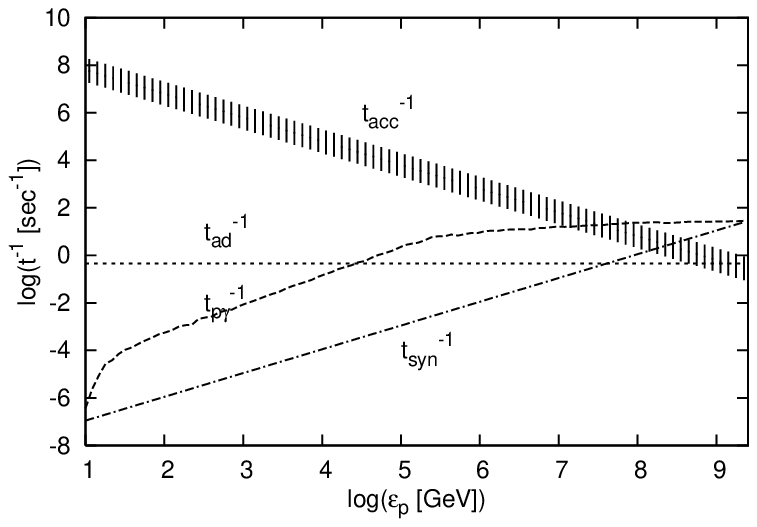}
\caption{\label{MAC1} Various cooling time scales and acceleration time scales for case A, $r=2 \times {10}^{13} \, \mr{cm}$, and $\epsilon _{B}=0.1$.
The hatched lines show an uncertainty of the acceleration time, which corresponds to $\eta=1 - 10$.}
\end{minipage}
\begin{minipage}{.02\linewidth}
\includegraphics[width=\linewidth]{white.eps}
\end{minipage}
\begin{minipage}{.48\linewidth}
\includegraphics[width=\linewidth]{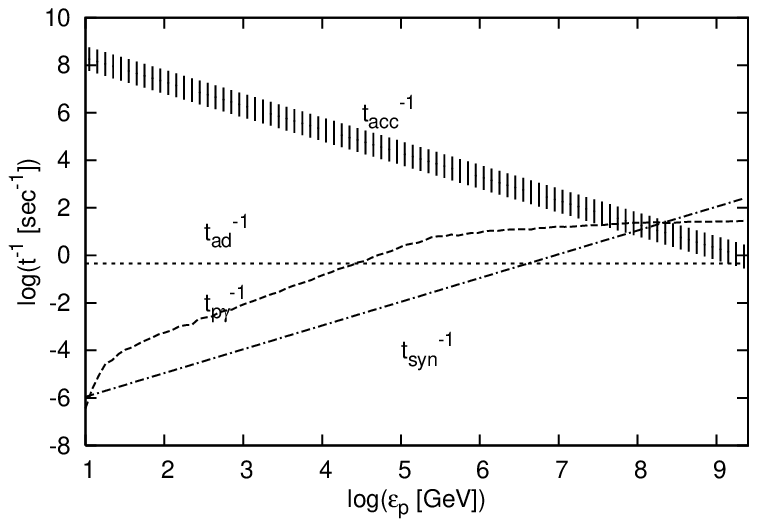}
\caption{\label{MAC2}  Various cooling time scales and acceleration time scales for case A, $r=2 \times {10}^{13} \, \mr{cm}$, and $\epsilon _{B}=1$.
The hatched lines show an uncertainty of the acceleration time, which corresponds to $\eta=1 - 10$.}
\end{minipage}
\end{figure*}

\section{\label{sec:level3}Results}
\subsection{\label{subsec:levelf}Photomeson Production Efficiency}
We calculate the proton cooling efficiency through photomeson production by the method explained above. The obtained results on $f_{p\gamma} \equiv 
t_{dyn}/t_{p\gamma}$ by using GEANT4 are shown in Fig. \ref{PC1}. For comparison, we also show the case where the cross section has a cutoff at 2 GeV in 
the proton-rest frame. In this case, we checked that our result agrees with Asano \cite{Asa1} within a few percent. Analytic rough approximation by 
$\Delta$-resonance is also shown in Fig. \ref{PC1} \cite{Wax3}. In this approximation we set $\sigma _{p\gamma} \approx 5 \times {10}^{-28} \, 
{\mr{cm}}^{2}$, $\bar{\varepsilon} \approx 0.3 \, \mr{GeV}$, $\Delta \bar{\varepsilon} \approx 0.3 \, \mr{GeV}$, and $\kappa _{p} \approx 0.2$ in the 
equation (\ref{pcool}). At $\gamma _p \alangle 10^{7} \mr{GeV}$, $\Delta$-resonance is a good approximation and the break energy is determined 
by $\varepsilon _{p}^{b} \varepsilon ^{b} \sim 0.3 \, {\mr{GeV}}^{2}$. Protons below the break energy mainly produce pions with photons whose energies 
are above the break energy, so it leads to $f_{\pi} \propto \varepsilon _{p}^{\beta-1}$. On the other hand, protons above the break energy mainly interact 
with harder photons, which leads to $f_{\pi} \propto \varepsilon _{p}^{\alpha-1}$. The photomeson production efficiency obtained by GEANT4 is larger than 
other two cases which have the cutoff and monotonically increasing in the high energy region of the order of $\gamma _p \arangle 10^{7} \mr{GeV}$. 
This is because $\kappa_{p} \approx 0.5-0.7$ rather than $\kappa _p \approx 0.2$ at $\Delta$-resonance are satisfied and multi-pion production occurs in 
this region. \\
\begin{figure*}[t]
\begin{minipage}{.48\linewidth}
\includegraphics[width=\linewidth]{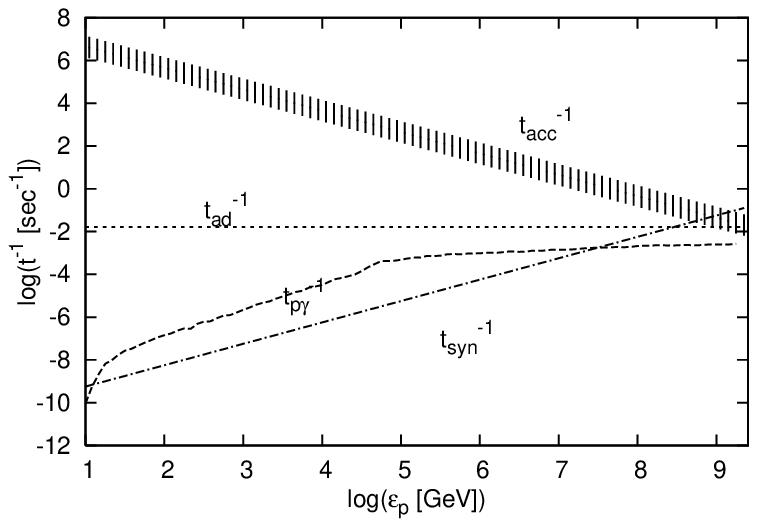}
\caption{\label{MAC3} The same as Fig. \ref{MAC1}. But a result for case B, $r=5.4 \times {10}^{14} \, \mr{cm}$, and $\epsilon _{B}=1$.}
\end{minipage}
\begin{minipage}{.02\linewidth}
\includegraphics[width=\linewidth]{white.eps}
\end{minipage}
\begin{minipage}{.48\linewidth}
\includegraphics[width=\linewidth]{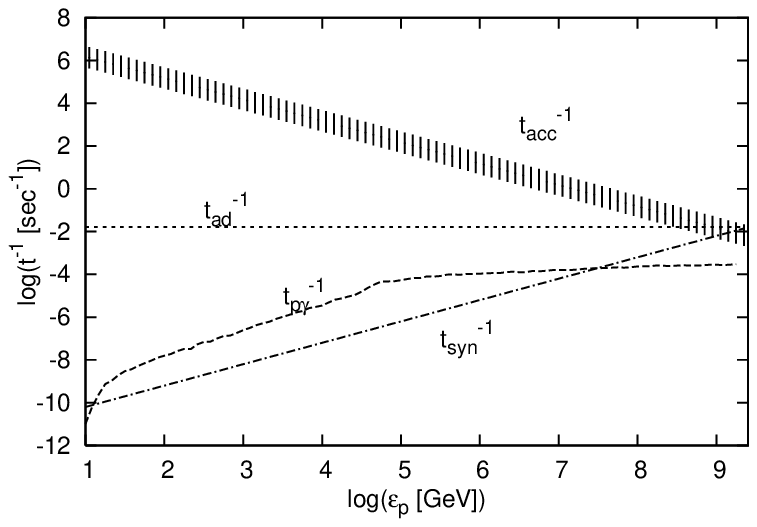}
\caption{\label{MAC4} The same as Fig. \ref{MAC1}. But a result for case B, $r=1.6 \times {10}^{15} \, \mr{cm}$, and $\epsilon _{B}=1$.}
\end{minipage}
\end{figure*}
In this paper we adopt three parameter sets for GRB isotropic energy, photon break energy, and photon spectral indices. That is, 
case A: $E_{\gamma}^{iso}=2 \times {10}^{51} \, \mr{ergs}$, $\varepsilon ^{b}=1 \, \mr{keV}$, and $\alpha=1, \, \beta=2.2$; case B: $E_{\gamma}^{iso}=2 
\times {10}^{52} \, \mr{ergs}$, $\varepsilon ^{b}=3 \, \mr{keV}$, and $\alpha=1, \, \beta=2.2$; case C: $E_{\gamma}^{iso}=2 \times {10}^{53} \, 
\mr{ergs}$, $\varepsilon ^{b}=6 \, \mr{keV}$, and $\alpha=0.5, \, \beta=1.5$. Fig. \ref{PC2} shows the proton cooling efficiencies for each case. Case A
 and case B have the fiducial spectral indices of GRB photons, while case C has a rather flatter photon spectrum. In addition, case C corresponds to the 
case where a more energetic burst will be observed.\\
Fig. \ref{MAC1} - Fig. \ref{MAC4} show various cooling time scales and acceleration time scale. Proton's maximum energy is determined by equation 
(\ref{mac}). Note that most of protons will be depleted in the case optically thick to photomeson production, even if a fraction of protons are 
accelerated up to the maximal energy indicated by the equation (\ref{mac}). For example, Fig. \ref{MAC1} and Fig. \ref{MAC2} corresponds to such cases. 
Only in the optically thin case, a significant fraction of protons with the maximum energy can escape from the source. In these figures, for comparison, 
we also show $\eta$ in the equation (\ref{acc}) as a parameter having the range of 1 - 10 (hatched lines). Fig. \ref{MAC1} shows the case where the 
photohadronic cooling is crucial for determining the maximum energy. Multiplicity and inelasticity will be important because the contribution of 
photohadronic cooling is comparable to synchrotron cooling if cross section has the cutoff at 2 GeV, or smaller than synchrotron cooling in the case of 
analytic $\Delta$-resonance approximation. At inner radii, photon density is large, so photohadronic process can be important unless $\epsilon _{B}$ is 
enough large for synchrotron cooling to be a dominant process. When the proton's maximum energy is determined by the photohadronic process, a significant 
fraction of protons with the very high energy cannot escape from the accelerating site in many cases of GRB parameters. It is only possible within the 
limited range of radii or for the case of the moderately smaller radiation energy than our cases. In many cases of possible parameter sets for GRBs, the 
synchrotron cooling is the most dominant process to determine the maximum energy and such a case is demonstrated by Fig. \ref{MAC2} and Fig. \ref{MAC3}. 
At outer radii, the dynamical time scale will be more and more important due to decreasing of photon energy density and magnetic energy density. Fig. 
\ref{MAC4} corresponds to such a case and the maximum energy of proton is restricted by $t_{ad}$. In our cases, significant acceleration of protons is 
possible only at larger radii, $r \arangle 10^{14} \, \mr{cm}$. This result is consistent with Asano \cite{Asa1}. Smaller $E_{\gamma}^{iso}$ and larger 
$r$ are favorable to generate UHECRs. Fig. \ref{MAC3} and Fig. \ref{MAC4} demonstrate such cases where sources are optically thin to photomeson production 
and the production of UHECRs is possible. The effects of multiplicity and high-inelasticity appear in the very high energy region and enhances 
photohadronic cooling of protons. However, these effects affect cooling time scale only for the cases of inner collision radii and smaller magnetic fields,
 $\epsilon _{B}=0.1$. In the next subsection, we will see these effects can become important only for limited cases.

\subsection{\label{subsec:levelg}Neutrino Spectrum and Flux}
\begin{figure*}[bt]
\begin{minipage}{.48\linewidth}
\includegraphics[width=\linewidth]{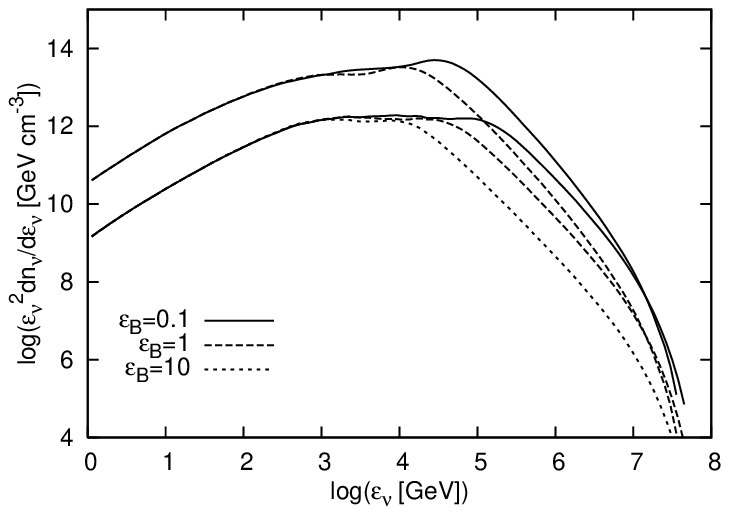}
\caption{\label{Gamma} Neutrino spectra in the comoving frame with $\epsilon _{B}$ changing for $r=2 \times {10}^{13} \, \mr{cm}$. The upper lines are for 
$\Gamma=100$ and $E_{\gamma}^{iso}=2 \times {10}^{52} \, \mr{ergs}$. The lower lines are for $\Gamma=300$ and $E_{\gamma}^{iso}=2 \times {10}^{51} \, 
\mr{ergs}$. Three cases are shown, $\epsilon _B=0.1$ (solid line), $\epsilon _B=1$ (dashed line), $\epsilon _B=10$ (dotted line).}
\end{minipage}
\begin{minipage}{.02\linewidth}
\includegraphics[width=\linewidth]{white.eps}
\end{minipage}
\begin{minipage}{.48\linewidth}
\includegraphics[width=\linewidth]{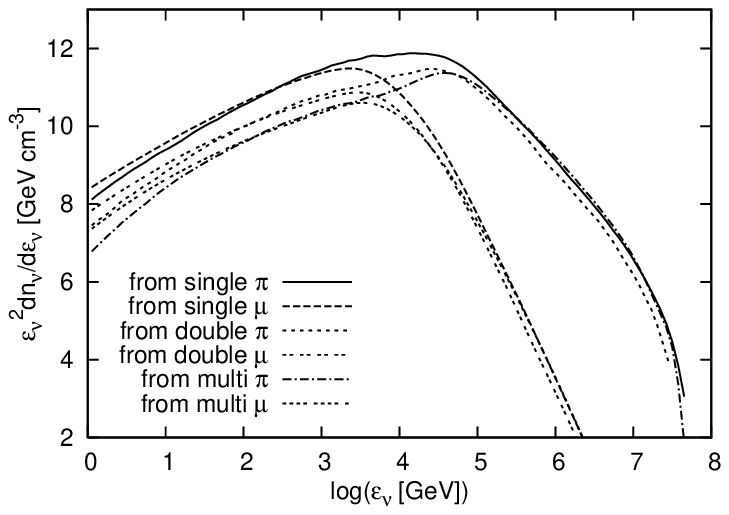}
\caption{\label{Mul1} Muon-neutrino (${\nu}_{\mu}+{\bar{\nu}}_{\mu}$) spectra in the comoving frame. Muon-neutrinos are produced by the decay of pion and 
muon which origins are single- or double- or multi-pion photomeson production. Each case is shown. A result for $r=2 \times {10}^{13} \, \mr{cm}$ on case A. 
$\epsilon _{B}$ is set to $1$.}
\end{minipage}
\end{figure*}
\begin{figure*}[tb]
\begin{minipage}{.48\linewidth}
\includegraphics[width=\linewidth]{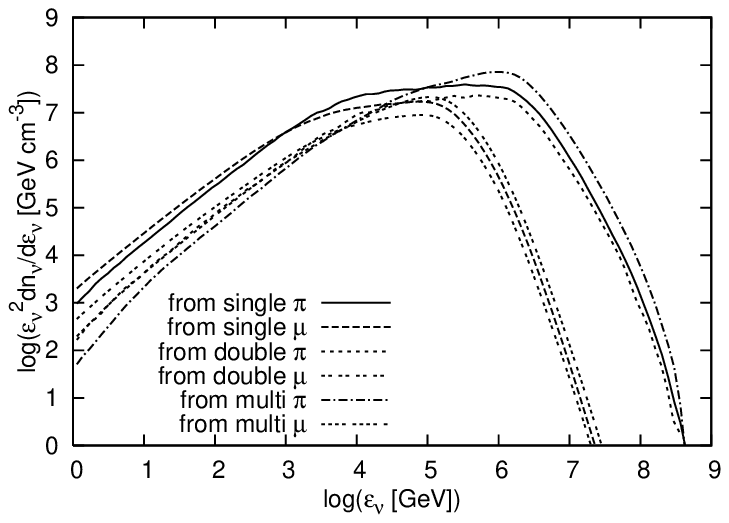}
\caption{\label{Mul3}  The same as Fig. \ref{Mul1}. But a result for $r=5.4 \times 10^{14} \, \mr{cm}$ on case B. $\epsilon _{B}$ is set to $1$.}
\end{minipage}
\begin{minipage}{.02\linewidth}
\includegraphics[width=\linewidth]{white.eps}
\end{minipage}
\begin{minipage}{.48\linewidth}
\includegraphics[width=\linewidth]{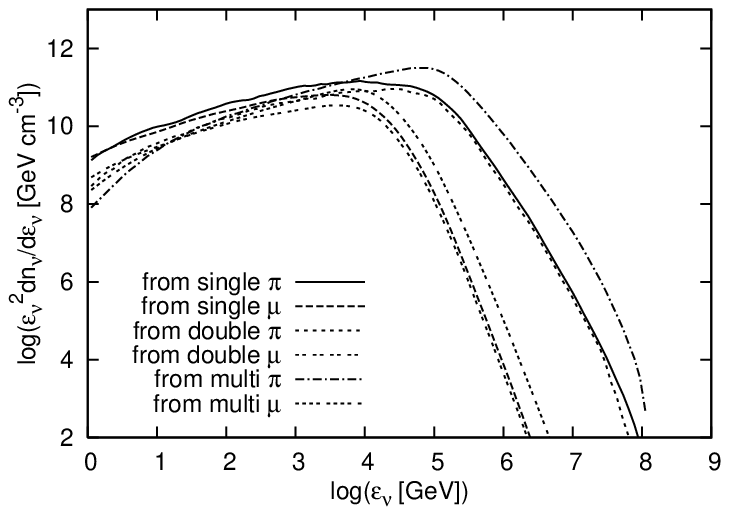}
\caption{\label{Mul4}  The same as Fig. \ref{Mul1}. But a result for $r=1.8 \times 10^{14} \, \mr{cm}$ on case C. $\epsilon _{B}$ is set to $0.1$.}
\end{minipage}
\end{figure*}
\begin{figure*}[tb]
\begin{minipage}{.48\linewidth}
\includegraphics[width=\linewidth]{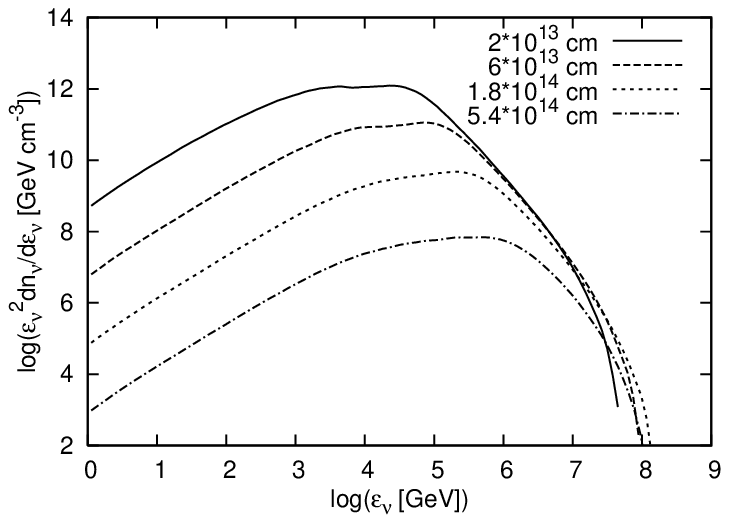}
\caption{\label{radii1} Muon-neutrino (${\nu}_{\mu}+{\bar{\nu}}_{\mu}$) spectra in the comoving frame for various collision radii on case A. The subshell 
width is $l=6.7 \times {10}^{10} \, \mr{cm}$. $\epsilon _{B}$ is set to $1$}
\end{minipage}
\begin{minipage}{.02\linewidth}
\includegraphics[width=\linewidth]{white.eps}
\end{minipage}
\begin{minipage}{.48\linewidth}
\includegraphics[width=\linewidth]{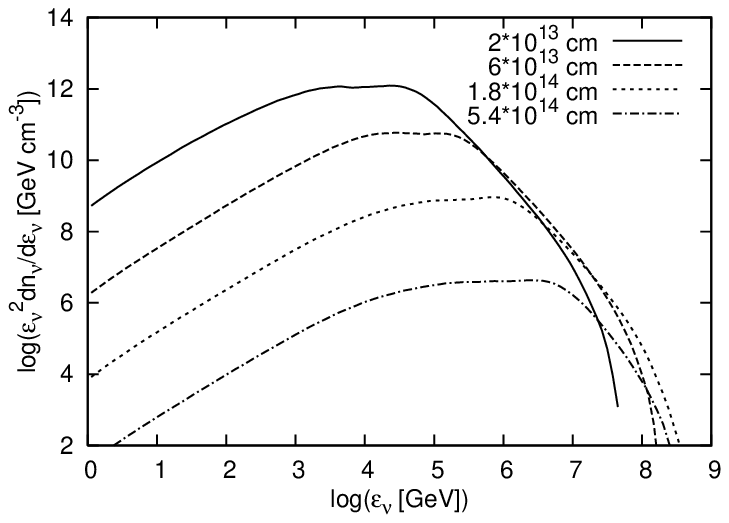}
\caption{\label{radii2} Muon-neutrino (${\nu}_{\mu}+{\bar{\nu}}_{\mu}$) spectra in the comoving frame for various collision radii on case A. The subshell 
width is given by $l=r/\Gamma$. $\epsilon _{B}$ is set to $1$}
\end{minipage}
\end{figure*}
We can get the spectrum of pions from photomeson production efficiencies calculated by GEANT4. From pion obtained spectra we can also calculate neutrino 
spectra following from the method explained in Sec. \ref{subsec:leveld} and \ref{subsec:levele}. Fig. \ref{Gamma} shows spectra of $\nu _{\mu}+
{\bar{\nu}}_{\nu}$ as one of our results for $r=2 \times {10}^{13} \, \mr{cm}$, $\Gamma=100$, and $\Gamma=300$. The high-energy break is mainly determined 
by $t_{syn}/t_{\pi}$. Hence the high-energy break changes satisfying $\varepsilon _{\nu}^{s} \propto {\epsilon}_{B}^{1/2}$ (see Fig. \ref{Gamma}). Note, 
from the equation (\ref{phden}), different values of $\Gamma$ give the similar neutrino spectra for the same $E_{\gamma}^{iso}$ and $r$ as long as 
$l=r/\Gamma$ is hold.\\
Fig. \ref{Mul1} - Fig. \ref{Mul4} show spectra from single-pion, double-pion, and multi-pion production origins. As seen in Fig. \ref{Mul1} and Fig. 
\ref{Mul3}, for the case of $\alpha=1$ and $\beta=2.2$ that is typical for GRB, the effects of double- and multi-pion are negligible or comparable to that 
of single-pion, whose contribution can be well described by $\Delta$-resonance approximation. Protons of energy $\varepsilon _{p} \alangle {10}^{5} \, 
\mr{GeV}$ can interact only with the steep part of the photon spectrum above the break, $dn/d\varepsilon \propto \varepsilon ^{-2.2}$. Hence the 
contribution of non-resonance, $\varepsilon _{p} \varepsilon \gg 0.16 \, \mr{GeV}$, is negligible. Protons with the energy $\varepsilon _{p} \arangle 
{10}^{5} \, \mr{GeV}$ can interact with the flatter part of the photon spectrum below the break, $dn/d\varepsilon \propto \varepsilon ^{-1}$. In this case, 
the contribution of double- and multi-pion production can be important because the flatter part of photon spectrum cover significant energy range. Such 
contribution can be crucial at the very high energy range, $\varepsilon _{p} \arangle {10}^{7} \, \mr{GeV}$. At inner radii the contributions of double- and 
multi-pion are negligible or comparable (see Fig. \ref{Mul1}). Even when it is comparable, such a region is above the high-energy break. Because the 
nonthermal proton's maximum energy is around $({10}^{8}-{10}^{8.5}) \, \mr{GeV}$, there are only a few very high energy protons which can produce 
multi-pions. At outer radii these contributions are comparable to or larger than single-pion fraction by a factor of $\sim (2-3)$ (see Fig. \ref{Mul3}). 
However, in most of the region where multi-pion production dominates, the resulting neutrino spectra are suppressed by synchrotron and adiabatic cooling 
processes because such a region belongs to the high energy region above around the high-energy break. For the case of $\alpha=0.5$ and $\beta=1.5$, which 
is the flatter photon spectrum, the multi-pion production has the significant effect for neutrino spectra. In this case there are sufficient high energy 
photons, which can interact with very high energy protons, $\varepsilon _{p} \arangle {10}^{7} \, \mr{GeV}$. Fig. \ref{Mul4} demonstrates one of such 
cases. In this case the contribution of multi-pion origin dominates single-pion origin by one order of magnitude even around the high-energy break. Note 
that in the very high energy region, not only inelasticity but also multiplicity are also high. As a result, the average pion's energy which can be estimated by the parent 
proton's energy multiplied by inelasticity and divided by multiplicity cannot be so large \cite{Muc2}.\\
The difference between neutrino spectra from pion decay and from muon decay is explained as follows. Neutrinos from muons dominate those emitted directly 
from pions in the low energy region. When a pion decays as ${\pi}^{\pm} \rightarrow {\mu}^{\pm}+{\nu}_{\mu}({\bar{\nu}}_{\mu})$, the energy fraction a muon 
obtains is typically $\sim m_{\mu}/m_{\pi} \sim 0.76$. Hence a direct neutrino from ${\pi}^{\pm}$ has a fraction of $\sim 0.24$. When a muon decays as
${\mu}^{\pm} \rightarrow e^{\pm}+{\nu}_{e}({\bar{\nu}}_{e})+{\nu}_{\mu}+{\bar{\nu}}_{\mu}$, each of three species will carry similar energy. But a neutrino 
from $\mu ^{\pm}$ has a smaller energy fraction than a direct neutrino from $\pi ^{\pm}$, since muon has the longer life time than pion so that it is 
subject to synchrotron cooling.\\
Fig. \ref{radii1} shows neutrino spectra that occur at various collision radii with the fixed shell width. In this case photon energy density changes 
with $U_{\gamma} \propto r^{-2}$. This result is consistent with Asano \cite{Asa1}. Fig. \ref{radii2} shows neutrino spectra which occurs at various 
collision radii with changing the shell width holding $l \approx r/\Gamma$. In this case photon energy density changes with $U_{\gamma} \propto r^{-2}l^{-1} 
\propto {\delta t}^{-3}$. Here, $\delta t$ is the typical variability time scale. Roughly speaking, the high-energy break is determined by $t_{syn}/t_{\pi}$
 again. This implies the high-energy break is proportional to $r{l}^{1/2}$. On the other hand the low-energy break is determined by the minimum energy of 
pions produced from protons. The dynamical time scale, $t_{dyn}$ is proportional to $l$, while $t_{p\gamma}$ and $t_{syn}$ are proportional to ${r}^{2}l$, 
so the proton cooling efficiencies, which is expressed by $f_{p\gamma} \equiv t_{dyn}/t_{p\gamma}$ and $f_{syn} \equiv t_{dyn}/t_{syn}$, are proportional 
to $r^{-2}$ independently of $l$ \cite{Asa1}. So, this minimum energy will depend on only $r$. Both the low-energy and high-energy break increase roughly 
proportionally to $r$. Fig. \ref{radii1} and Fig. \ref{radii2} confirm these statements.\\
\begin{figure*}[bt]
\begin{minipage}{.48\linewidth}
\includegraphics[width=\linewidth]{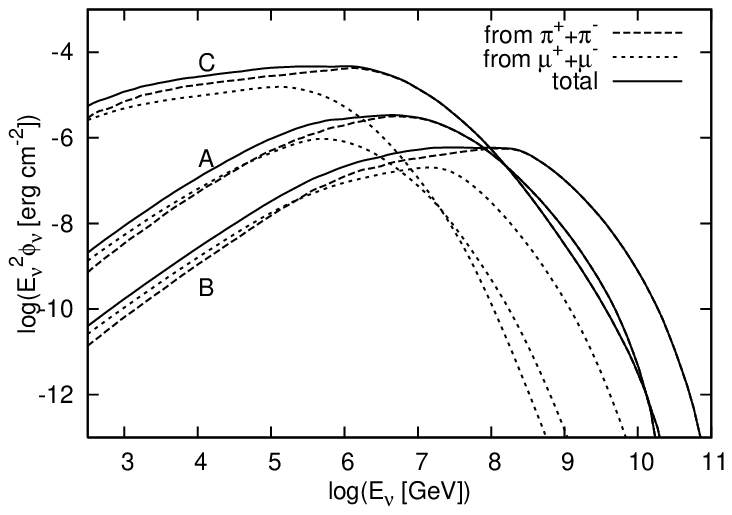}
\caption{\label{NBurst1} The observed muon-neutrino (${\nu}_{\mu}+{\bar{\nu}}_{\mu}$) spectra for one GRB burst at $z=1$. Neutrino spectra from pion and 
muon are shown respectively on set A with $N=200$, set B with $N=20$, set C with $N=20$. Set A and Set B are $\epsilon _{B}=1$, but Set C is 
$\epsilon _{B}=0.1$} 
\end{minipage}
\begin{minipage}{.02\linewidth}
\includegraphics[width=\linewidth]{white.eps}
\end{minipage}
\begin{minipage}{.48\linewidth}
\includegraphics[width=\linewidth]{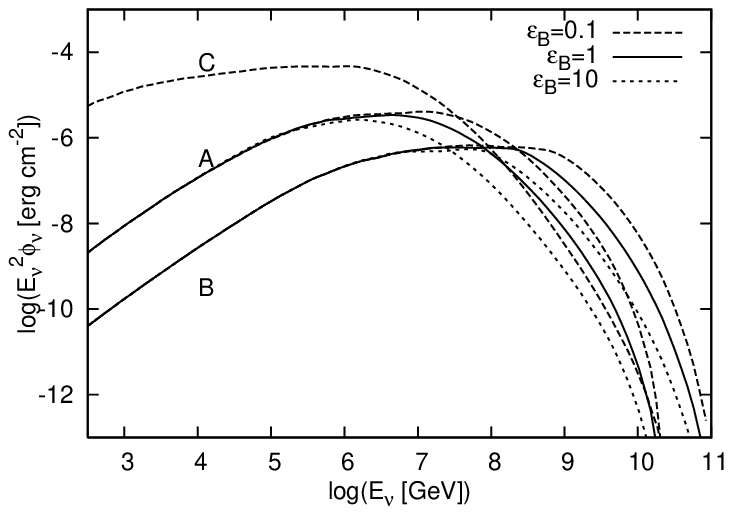}
\caption{\label{NBurst2} The observed muon-neutrino (${\nu}_{\mu}+{\bar{\nu}}_{\mu}$) spectra for one GRB burst at $z=1$. Neutrino spectra with various 
$\epsilon _{B}$ values are shown respectively on set A with $N=200$, set B with $N=20$, set C with $N=20$. Set A and Set B are $\epsilon _{B}=1$, but Set 
C is $\epsilon _{B}=0.1$}
\end{minipage}
\end{figure*}
To get the total neutrino spectrum of a single burst with a collision radius $r$ and a subshell width $l$ simply, we only have to multiply the number of 
collisions $N$. However, here we cumulate the neutrino spectra over various internal collision radii, assuming each collsion emits similar energy.
It is because internal shocks may occur within some range of distances, although the distribution of collision radii is not so clear. In the internal shock 
model, the sequence of collsions takes place. Collisions of the subshells with the smaller separation will occur at smaller radii. The subsequent collsions 
will occur at larger radii. For simplicity, we assume the number of such collisions is proportional to $\Delta/d$, where $\Delta$ is the total width 
of the shell. We also assume the fixed subshell width through one burst. We may need to take into account the spreading of subshells 
since internal shock radii are roughly comparable to the spreading radii of subshells \cite{Zha1}. Even so, it does not affect the resulting spectra so 
much because photon density becomes much smaller at larger radii. To improve our calculation, we will need a detailed calculation on the internal shock 
model \cite{Kob1,Kob2}, which is beyond the scope of this paper. We adopt three parameter sets in this study. That is, set A: 
$E_{\gamma}^{iso}=2 \times {10}^{51} \, \mr{ergs}$, $\varepsilon ^{b}=1 \, \mr{keV}$, $\alpha=1, \, \beta=2.2$, and $r = ({10}^{13} - 8.1 \times {10}^{14})
 \, \mr{cm}$, and $l=6.7 \times {10}^{10} \, \mr{cm}$; set B: $E_{\gamma}^{iso}=2 \times {10}^{52} \, \mr{ergs}$, $\varepsilon ^{b}=3 \, \mr{keV}$, 
$\alpha=1, \, \beta=2.2$, $r=(2.7 \times {10}^{14} - 7.3 \times {10}^{15}) \, \mr{cm}$, and $l=1.8 \times {10}^{12} \, \mr{cm}$; set C: $E_{\gamma}^{iso}=2
 \times {10}^{53} \, \mr{ergs}$, $\varepsilon ^{b}=6 \, \mr{keV}$, $\alpha=0.5, \, \beta=1.5$, $r=({10}^{13} - 8.1 \times {10}^{14}) \, \mr{cm}$, and 
$l=6.7 \times {10}^{10} \, \mr{cm}$. Set A demonstrates internal shocks begin at somewhat smaller radii, $r \sim {10}^{13} \, \mr{cm}$. In this set, the 
typical variability time scale is $\delta t \sim 30 \, \mr{ms}$ and if $N=200$, the typical jet angle is $\theta _{j} \sim 0.1 \, \mr{rad}$. Set B 
demonstrates internal shocks begin at $r \sim {10}^{14} \, \mr{cm}$. The typical variability time of this set is $\delta t \sim 0.3 \, \mr{s}$ and if 
$N=20$, the typical jet angle is $\theta _{j} \sim 0.1 \, \mr{rad}$. Set C is a very energetic case which has the flatter photon spectrum. Three sets are 
shown in Fig. \ref{NBurst1} and Fig. \ref{NBurst2}. To evaluate observed flux from one burst, we set a source at $z=1$. Obtained neutrino flux of set A and
 set B are comparable with Guetta et al. \cite{Gue1}. However, such levels of neutrino flux are hardly detected by ${\mr{km}}^{3}$ detector such as IceCube.
 Only the most powerful bursts or nearby sources can give a realistic chance for detection of $\nu _{\mu}$ \cite{Der1}. In our sets, only set C has the 
prospect for detection by IceCube. To see this, here we estimate neutrino events in IceCube. We use the following fitting formula of the probability of 
detecting muon neutrinos \cite{Iok1,Raz1}.
\begin{equation}
P(E_{\nu})=7 \times 10^{-5} {\left( \frac{E_{\nu}}{{10}^{4.5} \, \mr{GeV}}\right)}^{\beta}
\end{equation} 
where $\beta=1.35$ for $E_{\nu}<{10}^{4.5} \, \mr{GeV}$, while $\beta=0.55$ for $E_{\nu}>{10}^{4.5} \, \mr{GeV}$. Using a geometrical detector area of 
$A_{det}=1 {\mr{km}}^{2}$, the numbers of muon events from muon-neutrinos a burst are given by,
\begin{equation}
N_{\mu}(>E_{\nu, 3})=A_{det} \int _{\mr{TeV}} \!\!\!\!\!\!\! dE_{\nu} \, P(E_{\nu}) \, \frac{dN_{\nu}(E_{\nu})}{dE_{\nu}dA}
\end{equation}
where $E_{\nu}=10^{3} \, \mr{GeV} \, E_{\nu,3}$. Hence, the numbers of muon-neutrinos to be expected by IceCube for set C with $N=20$ are $N_{\mu}=1.9$ 
particles. In the case of set A with $N=200$ and set B with $N=20$, we obtain $N_{\mu}=0.05$ and $N_{\mu}=0.004$ respectively. Of course, if 
$\epsilon _{acc}$ is more larger, flux can be enhanced. But too large $\epsilon _{acc}$ will be suspicious. We will discuss this later.\\
\begin{figure*}[bt]
\begin{minipage}{.48\linewidth}
\includegraphics[width=\linewidth]{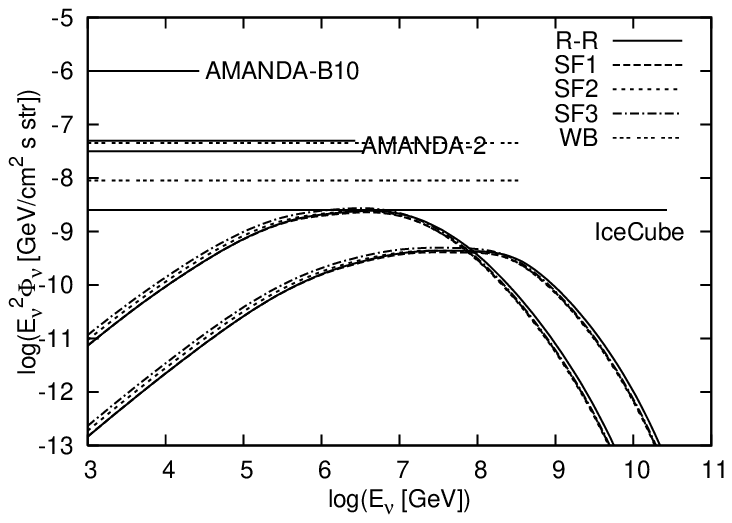}
\caption{\label{NBG1} Diffuse neutrino background from GRBs for $z_{\mr{max}}=7$ on several SFR models. R-R means the case using Rowan-Robinson SFR. SF1 - 
SF3 correspond to the models of Porciani \& Madau. The left lines are for set A, while the right is for set B. For comparison, we show WB bounds (three 
dashed lines). The upper WB bound is for z-evolution of QSOs. The lower is for no z-evolution. $\epsilon _{acc}=10$ and $\epsilon _{B}=1$ are assumed.}
\end{minipage}
\begin{minipage}{.02\linewidth}
\includegraphics[width=\linewidth]{white.eps}
\end{minipage}
\begin{minipage}{.48\linewidth}
\includegraphics[width=\linewidth]{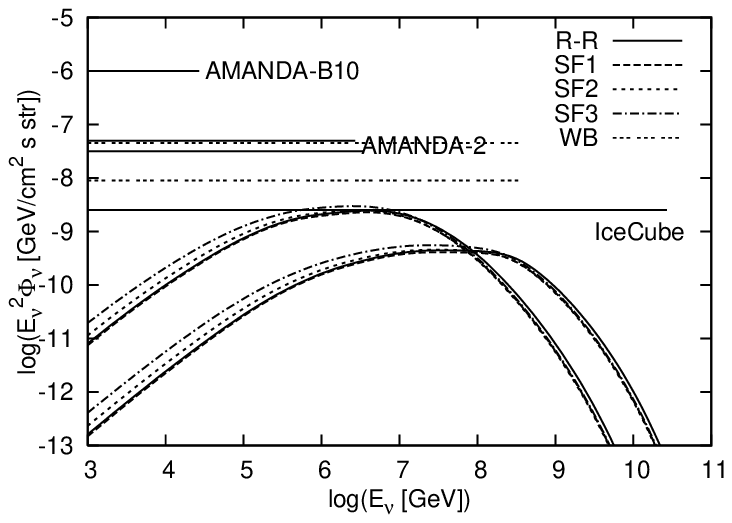}
\caption{\label{NBG2} Diffuse neutrino background from GRBs for $z_{\mr{max}}=20$ on several SFR models. R-R means the case using Rowan-Robinson SFR. SF1 - 
SF3 correspond to the models of Porciani \& Madau. The left lines are for set A, while the right is for set B. For comparison, we show WB bounds (three
dashed lines). The upper WB bound is for z-evolution of QSOs. The lower is for no z-evolution. $\epsilon _{acc}=10$ and $\epsilon _{B}=1$ are assumed.}
\end{minipage}
\end{figure*}
\begin{figure}[bt]
\includegraphics[width=\linewidth]{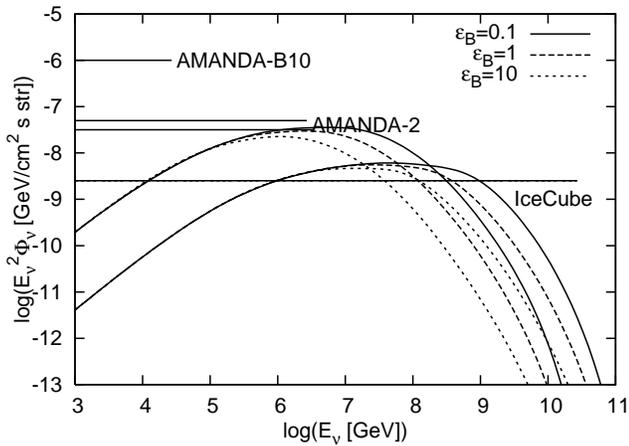}
\caption{\label{NBG3} Diffuse neutrino background from GRBs for $z_{\mr{max}}=20$ with $\epsilon _{B}$ changing. But the large baryon-loading factor, 
$\epsilon _{acc}=100$, is assumed. The upper lines are for set A, while the lower lines are for set B. All lines use the SF3 model.}
\end{figure}
Since we obtain neutrino spectra on various parameters above, we can calculate a diffuse neutrino background from GRBs for our specific parameter sets. 
The results are shown in Fig. \ref{NBG1} - Fig. \ref{NBG3}. Fig. \ref{NBG1} is for $z_{\mr{max}}=7$. Fig. \ref{NBG2} and Fig. \ref{NBG3} are for 
$z_{\mr{max}}=20$. Our adopted SFR models generate similar results. When $z_{\mr{max}}$ is $20$, SF3 model gives higher flux by a factor of $\sim (2-3)$ 
below the low-energy break than other SFR models. This is because SF3 model predicts higher SFR at high redshifts. Whether $z_{\mr{max}}$ is $7$ or $20$ 
does not affect neutrino flux up to a factor. Neutrino signals from GRBs can be marginally detected or not by IceCube in both figures. In fact, when
 $\epsilon _{B}$ is set to $1$ and we use SF3 model, we can obtain $N_{\mu}=14$ particles a year for $z_{\mr{max}}=7$ and $N_{\mu}=17$ particles a year for
 $z_{\mr{max}}=20$. On the other hand, in the case of set B, we get $N_{\mu}=1.2$ particles a year for $z_{\mr{max}}=7$ and $N_{\mu}=1.5$ particles a year 
for $z_{\mr{max}}=20$. \\
Our result for set A is very similar to the prediction of Waxman \& Bahcall \cite{Wax2}, although our ground is different from theirs. In this parameter 
set, our calculation on neutrino background from GRBs also satisfy WB bound, although this case is optically thick to photomeson production at inner radii, 
$r \alangle {10}^{14} \, \mr{cm}$. Set B expresses neutrino spectra for the case of larger collision radii than set A. In this parameter set, GRB sources 
are optically thin to photomeson production and can be sources of very high energy cosmic rays. It has to satisfy WB bound and this implies $\epsilon _{acc}$ 
can be constrained from UHECR observations if set B is fiducial for GRBs and other parameters are appropriate. If GRBs are sources of UHECRs, this set 
implies GRB neutrino fluxes with z-evolution, $E_{\nu}^{2} dF_{\nu}/dE_{\nu}d\Omega \sim {10}^{-8} \, \mr{GeV \, {cm}^{2} \, {s}^{-1} \, {str}^{-1}}$. 
However, this suggests very large baryon-loading factor, $\epsilon_{acc} \sim 100$ if current GRB rate estimation is correct and our model is valid. If set 
A is more fiducial, similar arguments leads to even larger baryon-loading factor because UHECRs can be accelerated only at large radii. So GRBs are not main 
sources of UHECRs in set A. If we adopt larger isotropic energy by one order in this set with $N$ fixed, the flux level will increase within a factor, 
because larger isotropic energy implies smaller $\theta _{j}$ (but for set B, it will increase by one order of magnitude, which is easily seen from the 
equations (\ref{phden}), (\ref{pcool}), and (\ref{getpion})). Fig. \ref{NBG3} shows $\epsilon _{B}$ dependence of neutrino background. But the near extreme
 case, $\epsilon _{acc}=100$ is presumed. If it is possible, neutrino will be surely observed by the detector such as IceCube. For example, when 
$\epsilon _{B}$ is set to $1$ and we use SF3 model, we obtain $N_{\mu}=170$ particles a year for set A with $z_{\mr{max}}=20$ and $N_{\mu}=15$ particles a 
year for set B with $z_{\mr{max}}=20$. Observations of neutrino background, if detected, may give us the evidence of protons being accelerated in GRBs, 
support to the internal shock model of GRBs, and information about these parameters independently of X/$\gamma$ rays from GRBs.\\
Finally, we summarize the parameter dependence of our results as follows. The collision radii, $r$ and the width of a subshell, $l$ determine the photon 
energy density. The larger radii and width of a subshell make the resulting neutrino emissivity smaller. The low-energy break of neutrino spectrum, 
$\varepsilon _{\nu}^{b}$ is determined by the photon break energy, but it is also roughly proportional to $r$. The high-energy break is determined by the 
synchrotron (or adiabatic) cooling and satisfies $\varepsilon _{\nu}^{s} \propto {\epsilon _{B}}^{1/2}$. We take $\epsilon _{B}=0.1, 1, 10$ (in the case of 
aftergrow, $\epsilon _{B}=0.1$ is preferred). The high-energy break is also roughly proportional to $r^{2}l$. One of the most important parameters is the 
nonthermal baryon-loading factor, $\epsilon _{acc}$. We set $\epsilon _{acc}$ to $10$ except Fig. \ref{NBG3} and the larger $\epsilon _{acc}$ can raise 
the flux level of neutrino, although too large $\epsilon _{acc}$ is not plausible. We take $E_{\gamma,tot}^{iso} \sim ({10}^{52}-{10}^{54}) \, \mr{ergs}$, 
and more energetic bursts can produce more neutrinos when other parameters are fixed. 

\section{\label{sec:level4}Summary and Discussion}
In this paper we calculate proton's photomeson cooling efficiency and resulting neutrino spectra from GRBs quantitatively. We are able to include 
pion-multiplicity and proton-inelasticity by executing GEANT4. These effects of multi-pion production and high-inelasticity in the high energy region 
enhance the proton cooling efficiency in this region, so they help prevent protons from accelerating up to the ultra-high energy region in several cases. 
But in many cases, the synchrotron loss time scale and the dynamical time scale determine proton's maximum energy. Furthermore, resulting neutrino spectra 
are not so sensitive to the nonthermal proton's maximum energy except the very high-energy region above the high-energy break.\\
In our cases GRBs are optically thick to photomeson production at $r \alangle {10}^{14} \, \mr{cm}$, 
while at $r \arangle {10}^{14} \, \mr{cm}$ GRBs are optically thin to it, so the production of UHECRs is possible at larger radii. This result is 
consistent with Asano \cite{Asa1}. Using the obtained proton cooling efficiency, we can calculate neutrino spectra. The effects from multi-pion production 
on resulting spectra are also calculated. We show that the contribution of multi-pion is almost negligible at inner radii but can be larger than that of 
single-pion at outer radii by a factor. In addition, the contribution of multi-pion production is somewhat sensitive to the proton's maximum energy, which 
would be actually difficult to determine precisely due to uncertainty of $\eta$. But such a contribution can be significant for the flatter photon spectrum,
 even though mesons lose their energy through these cooling process. Radiative cooling of pions and muons plays a crucial role in resulting spectra. 
Neutrino spectra are suppressed above the high-energy break energy. If the magnetic field is strong, such suppression becomes large and vice versa, if the 
magnetic field is weak, such suppression becomes weak. The observations of neutrino has the possibility that gives us some information about such a 
parameter of GRBs independently of gamma-ray observation. However, as shown in Dermer \& Atoyan \cite{Der1}, only the most powerful bursts, which are 
brighter than $\sim {10}^{53} \, \mr{ergs}$ or bursts at $z \alangle 0.1$, produce detectable neutrino bursts with a ${\mr{km}}^{3}$ detector such as 
IceCube. Set C of our parameter sets would be one of such detectable cases, but only a few neutrinos are expected even for the brightest bursts. For one 
neutrino burst, we adopt $\Gamma=300$ and consider only the on-axis observations. If we observe at off-axis, we will observe lower energy neutrinos.\\
A diffuse neutrino background is also calculated in this paper. In our specific parameter sets, neutrino background observations by IceCube can expect a 
few or a few tens order of neutrinos per year, although it is important which parameter set is fiducial. Extrapolation of SFR to high redshifts may not be 
valid. Even so, our results would not be so much affected as we have seen above. In addition, GRBs may trace not SFR, but metallicity. In the collapsar 
model, the presence of a strong stellar wind (a consequence of high metallicity) would hinder the production of a GRB, therefore metal-poor hosts would be 
favored sites \cite{Mac1}. There remains large uncertanty at low metallicity at present, and as more bursts are followed up and their environments are 
better studied by Swift, this correlation will be testable. However, our results will not be changed so much even when we can take into account this.\\
Throughout this paper, we set $N$ to the range of $\sim (10-100)$. If we change $N$, a neutrino signal from one burst will change 
with $N$. On the other hand, when we calculate a neutrino background, the results are not changed unless we fix $E_{\gamma}^{iso}$. This is because 
$\theta _{j}$ varies according to the change of $N$ since we fix $E_{\gamma}^{iso}$ and $E_{\gamma, tot}$. However, the results will be changed when 
we fix $\theta _{j}$ and $E_{\gamma, tot}$. For example, on set B, we will get the lower flux level by one order of magnitude if we adopt $N=200$ fixing 
$E_{\gamma,tot}$, $\theta _{j}$, and $\epsilon _{acc}$. This is easily seen from the $E_{\gamma,tot}^{iso}=f_b^{-1}E_{\gamma,tot}$, and the equations 
(\ref{phden}), (\ref{pcool}), and (\ref{getpion}). When we fix $E_{\gamma,tot}$, $\theta _{j}$, and $\epsilon _{acc}$, more shells mean lower photon 
energy density. For set B, which is the case optically thin to photomeson production, it leads to decrease $dn_{\pi}/d\varepsilon _{\pi}$ roughly by two 
orders of magnitude. Hence, the total neutrino spectra will be lower by one order of magnitude. If we fix $\theta _{j}$, our results in which $N$ is set to
 $\sim 10$, would give the reasonable flux level in the optimistic case.\\
In our parameter sets both one burst emission and a diffuse neutrino background give the neutrino flux we can barely observe by IceCube. To raise flux, 
GRBs require the larger nonthermal baryon-loading factor, $\epsilon _{acc}$, which is difficult to estimate from microphysics at present. However, there are 
some clues and assuming too large $\epsilon _{acc}$ will not be plausible. First, as seen in Sec. \ref{subsec:levelg}, UHECRs observations can give the 
upper limit to $\epsilon _{acc}$. Furthermore, the large baryon-loading factor suggests a significant contribution of the accelerated protons in the 
observed hard radiation through secondaries produced in photomeson production. Such emission is expected to be observed in the multi-GeV energy range by 
electromagnetic cascades \cite{Wax3,Bot1,Wic1}. If the flux level of multi-GeV emission is comparable to the neutrino flux obtained in this paper, the flux 
level will be below that of X/$\gamma$ emission and the EGRET limit. But assumed photon spectra will be modified by the radiation from secondaries. In 
addtion, such high energy emission may be detected by the near future GLAST observation \cite{Mce1}. Although it is important both to compare X/$\gamma$ 
emission with such multi-GeV emission and to investigate whether GLAST can detect such multi-GeV emission or not, a detailed calculation for this purpose 
is needed and it is beyond the scope of this paper. If such a calculation is done, GLAST observation in the near future has the possibility to give more 
information about $\epsilon _{acc}$. This will be the second clue about $\epsilon _{acc}$. Third, $\epsilon _{acc}$ will be constrained by the GRB total 
explosion energy, which is still unknown. The large baryon-loading factor leads to the large explosion energy. For example, if the true total explosion 
energy is $\alangle 10^{53} \, \mr{ergs}$, this suggests $\epsilon _{acc} \alangle 100$.\\
On the other hand, GRBs associated with supernovae may imply the isotropic kinetic energy, $E_{kin}^{iso} \arangle {10}^{52}
 \, \mr{ergs}$, which is larger than usual supernovae by one order of magnitude \cite{Nom1}. This leads to a collapsar model as a failed supernova in the 
sense that a core collapse event failed to form a neutron star and instead produced a black hole \cite{Mac1}. Here, we assume the internal shock scenario 
is correct and a collapsar model is valid. If the true total explosion energy of a collapsar is assumed to be $\alangle 10^{53} \, \mr{ergs}$, neutrino 
spectra will be limited by $f_{cl} \epsilon _{acc} \alangle {10}^{-1}$ because $f_{cl}$ is constrained by $f_{cl} \alangle {10}^{-3}$ from current GRB rate 
estimations. So if observed neutrino spectra is higher than these values, the collapsar model cannot explain the spectra alone. On the other hand, the 
supernova model predicts higher neutrino flux by a pulsar wind \cite{Kon1,Der1}. If the observed neutrino flux is higher than expected in the collapsar 
model, the supernova model might be likely to exist.\\
If IceCube can detect neutrinos and confirms they have the expected level of neutrino flux by our calculation, this will be one of the evidences that a 
significant fraction of protons can be accelerated and our employed internal shock model is valid. Of course, to estimate neutrino background more 
precisely and make our discussion justified, we have to choose the most fiducial parameters for GRBs. The results depend on photon density, and the 
contribution from a fraction of bursts with large photon density might be large. So, we should take into account the respective distributions of parameters 
to execute the most refined calculaion. Unfortunately, many parameters have large uncertainty at present. For this reason, we calculate for a wide range of 
these parameters in this paper. The signature of GRBs may depend on $z$. For example, the total isotropic energy of a GRB and photon spectral indices may 
depend on $z$. More and more observations in the near future and more refined theoretical models will allow our results to be improved.\\
So far, we have not taken account of neutrino oscillations. Since we have considered many decaying modes, the production ratio of high energy muon 
and electron neutrinos is not 2:1 exactly. However, the neutrinos will be almost equally distributed among flavors as a result of vacuum neutrino 
oscillations \cite{Wax3}. So there may be a possibility that tau neutrinos are detected through double bang events \cite{Ath1}.\\
In summary, we obtain the neutrino spectrum from GRBs quantitatively by using GEANT4 simulation kit. We show that photomeson cooling process can constrain 
the proton's maximum energy and the effects of multi-pion production and high-inelasticity can enhance the cooling efficiency. Furthermore, these effects 
affect the resulting neutrino spectra slightly and can be significant for the flatter photon spectrum. We quantitatively checked radiative cooling of pion 
and muon play a crucial role, which is controlled by $\epsilon _{B}$. We also confirmed that UHECRs can be accelerated at $r \arangle {10}^{14} \, 
\mr{cm}$. We have calculated not only neutrino spectra from one burst but also the GRB diffuse neutrino background using several SFR models. We have found 
our specific parameter sets give neutrino spectra comparable with the prediction of Waxman \& Bahcall \cite{Wax3} without supposing the very large 
nonthermal baryon-loading factor, which is necessary for the assumption that GRBs are main sources of UHECRs. We have also discussed influences on neutrino 
spectra by changing parameters. Such a study is important since there are many parameters. If neutrino signals are detected by AMANDA, ANTARES, NESTOR, or 
IceCube, it will be one of the evidences that protons can be accelerated to very high energy in GRBs and the internal shock model of GRBs are plausible. 
Furthermore, such neutrino observations in the near future may give us some information about the nonthermal baryon-loading factor and the inner engine of 
GRBs.    

\begin{acknowledgments}
We are very grateful to K. Asano for many useful comments and advises for calculation. We also thank the referee and K. Ioka for a lot of profitable 
suggestions. K.M. is grateful to K. Miwa and G. Normann for using GEANT4 simulation. We also thank M. Mori, H. Kurashige, T. Hyodo, and D. Wright for 
GEANT4. This work is in part supported by a Grant-in-Aid for the 21st Century COE ``Center for Diversity and Universality in Physics'' from the Ministry 
of Education, Culture, Sports, Science and Technology of Japan. S.N. is partially supported by Grants-in-Aid for Scientific Research from the Ministry of 
Education, Culture, Sports, Science and Technology of Japan through No. 14102004, 14079202, and 16740134.
\end{acknowledgments}

\appendix

\newpage 

\end{document}